\documentclass{article}
    \PassOptionsToPackage{numbers, compress}{natbib}

\usepackage{neurips_data_2022}

\usepackage[utf8]{inputenc} 
\usepackage[T1]{fontenc}    
\usepackage{hyperref}       
\usepackage{url}            
\usepackage{booktabs}       
\usepackage{amsfonts}       
\usepackage{nicefrac}       
\usepackage{microtype}      
\usepackage{xcolor}         

\usepackage{graphicx}
\usepackage{amsmath}
\usepackage{amssymb}
\usepackage{booktabs}

\usepackage{times}

\usepackage{soul}
\usepackage{url}
\usepackage{hyperref}
\usepackage{graphicx,xfp}
\usepackage{multicol}
\usepackage{multirow}
\usepackage{color}
\usepackage{xcolor}
\usepackage{booktabs}
\usepackage{graphicx}
\usepackage{nicefrac}
\usepackage{amsmath}
\usepackage{array}
\newcolumntype{P}[1]{>{\centering\arraybackslash}p{#1}}

\usepackage{graphicx}
\usepackage{fancyhdr}
\usepackage{amssymb}
\usepackage{url}
\usepackage{color}
\usepackage{tikz}
\usetikzlibrary{positioning}
\usepackage{pgfplots}
\usepackage{marvosym}

\usepackage{float}
\usepackage{todonotes}

\usepackage{subcaption}

\usepackage{hyperref}

\usepackage{tikz}
\usepackage{tikzscale}
\usepackage{xspace}
\usepackage{pgfplots}
\usepgfplotslibrary{groupplots}
\pgfplotsset{compat=1.16}
\usepackage{xparse}
\NewDocumentCommand{\Log}{o}{%
  \IfNoValueTF{#1}{}{{}^{#1}\!}\log}%

\usepackage{algorithm}
\usepackage[noend]{algpseudocode}

\usetikzlibrary{external}
\usepackage{color}

\usepackage{cite}
\usepackage{amsmath,amssymb,amsfonts}

\usepackage{graphicx}
\usepackage{textcomp}

\newtheorem{definition}{Definition}[section]

\newcommand{\ie}{\emph{i.e.}}
\newcommand{\eg}{\emph{e.g.}}

\usepackage[utf8]{inputenc} 
\usepackage[T1]{fontenc}    
\usepackage{hyperref}       
\usepackage{url}            
\usepackage{booktabs}       
\usepackage{amsfonts}       
\usepackage{nicefrac}       
\usepackage{microtype}      

\title{Benchmarking Machine Learning Robustness in Covid-19 Genome Sequence Classification}

\author{%
  Sarwan Ali \\
  Department of Computer Science\\
  Georgia State University\\
  Atlanta, GA 30303 \\
  \texttt{sali85@student.gsu.edu} \\
  \And
  Bikram Sahoo \\
  Department of Computer Science\\
  Georgia State University\\
  Atlanta, GA 30303 \\
  \texttt{bsahoo1@student.gsu.edu} \\
  \And
  Alexander Zelikovskiy \\
  Department of Computer Science\\
  Georgia State University\\
  Atlanta, GA 30303 \\
  \texttt{alexz@gsu.edu} \\
  \And
  Pin-Yu Chen\\
    AI Foundations Group\\
IBM T. J. Watson Research Center\\
Yorktown Heights, NY\\
  \texttt{pin-yu.chen@ibm.com} \\
  \And
  Murray Patterson \\
  Department of Computer Science\\
  Georgia State University\\
  Atlanta, GA 30303 \\
  \texttt{mpatterson30@gsu.edu} \\
}

\begin{document}

\maketitle

\begin{abstract}
The rapid spread of the COVID-19 pandemic has resulted in an unprecedented amount of sequence data of the SARS-CoV-2  genome
--- millions of sequences and counting.  This amount of data, while
being orders of magnitude beyond the capacity of traditional
approaches to understanding the diversity, dynamics, and evolution of
viruses is nonetheless a rich resource for machine learning (ML)  approaches as alternatives for extracting such
important information from these data.  It is of hence utmost
importance to design a framework for testing and benchmarking the
robustness of these ML models.

This paper makes the first effort (to our knowledge) to benchmark the robustness of ML models by simulating  biological sequences with errors.
In this paper, we introduce several ways to perturb SARS-CoV-2 genome sequences to mimic the error profiles of common
sequencing platforms such as Illumina and PacBio.  We show from
experiments on a wide array of ML models that some simulation-based approaches are more robust (and accurate) than others for specific embedding methods 
to certain adversarial attacks to the input sequences. 
Our benchmarking framework may assist researchers in properly assessing different ML models and help them understand the behavior of the SARS-CoV-2 virus or avoid possible future pandemics.
\end{abstract}

\section{Introduction}
\label{sec:intro}
The novel (RNA) coronavirus, SARS-CoV-2, was identified in January 2020~\citep{wu2020new}, which began 
the COVID-19 pandemic that is still ongoing today.  With the help of
sequencing technology and phylogenetic analysis, the scientific
community disclosed that this novel coronavirus has 50\% similarity
with the Middle-Eastern Respiratory Syndrome Coronavirus (MERS-CoV),
79\% sequencing similarity to Severe Acute Respiratory Syndrome
Coronavirus (SARS-CoV) --- also known simply as ``SARS'' --- and more
than 85\% similarity with a coronavirus found in bats.  Further
studies confirmed that bats are the likely reservoir of these
coronaviruses; however, the ecological separation of bats from humans
indicates that some other organisms may have acted as intermediate
hosts.  Considering all scientific evidence, the International
Committee on Taxonomy of Viruses named the novel RNA virus
SARS-CoV-2~\citep{wu2020new,park2020epidemiology,zhang2020genomic}.

\begin{figure}[h!]
  \centering
  \includegraphics[scale = 0.25] {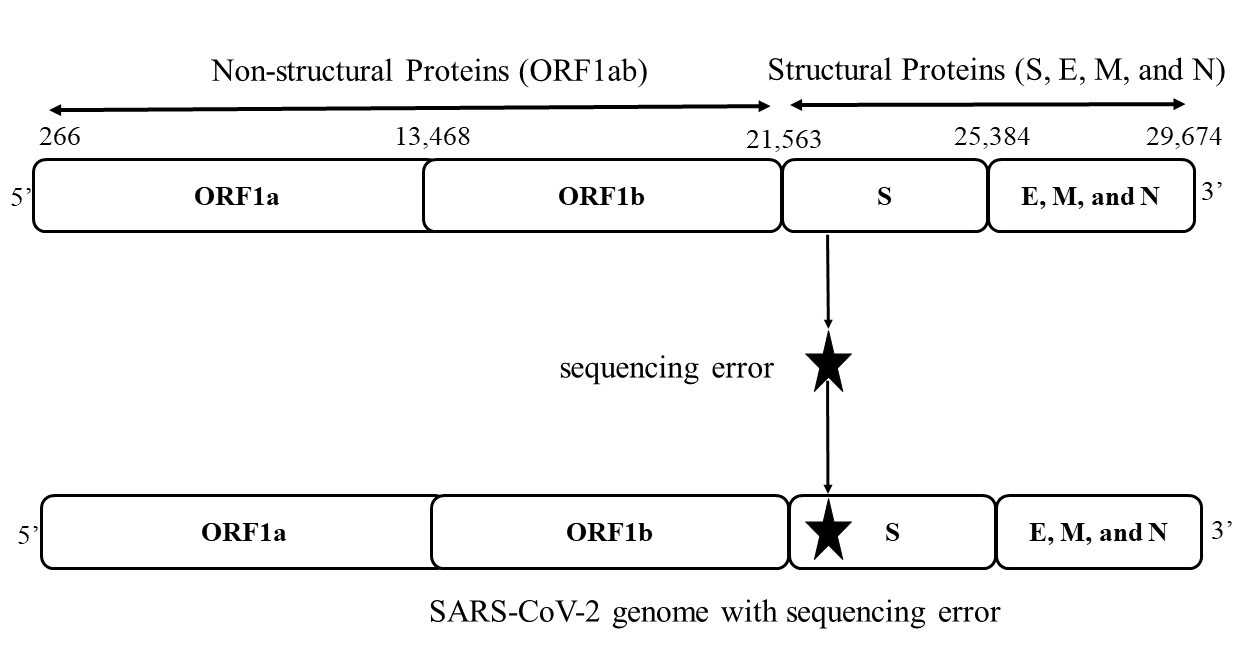}
  \caption{The SARS-CoV-2 genome codes for several proteins, including
    the surface, or spike (S) protein, where mutations happen disproportionately often~\citep{CDS_variantDef,kuzmin2020machine}.  Sequencing errors can bias the identification of a variant~\citep{huang2020structural,arons2020presymptomatic}.}
  \label{fig_spike_seq}
\end{figure}

RNA viruses generally introduce errors during replication, the
resulting mutations incorporated into the viral genome after repeated
replication within a single host, generating a heterogeneous population
of viral quasi-species.  However, SARS-CoV-2 has an excellent
proofreading mechanism that encodes a nonstructural protein 14 (nsp14)
allowing it to have a $10$-fold lower mutation rate than  typical RNA
viruses.  Epidemiologists estimate that SARS-CoV-2 undergoes $33$ genomic
mutations per year on average.  Some of these mutations are
advantageous, leading to more infectious variants of SARS-CoV-2
that continue to emerge~\citep{nelson2021tracking}.  
Moreover, each major variant can be characterized or differentiated by a
handful of mutations~\citep{CDS_variantDef}.
Hence, a sequencing error in the SARS-CoV-2 genome (see Figure~\ref{fig_spike_seq}) may lead to a false variant and influence the current study of SARS-CoV-2 virus~\citep{CDS_variantDef,kuzmin2020machine}.

The diminishing cost of next-generation sequencing (NGS) technology
has aided scientists from different parts of the world to generate
large volumes of SARS-CoV-2 whole-genome sequencing (WGS) data.  The Centers for Disease
Control and Prevention (CDC) of the United States has also provided a
wealth of information on resources, tools, and protocols for
SARS-CoV-2 WGS data from different sequencing platforms such as Illumina,
PacBio, and Ion Torrent.  Finally, the Global Initiative on Sharing
All Influenza Data (GISAID) hosts the largest SARS-CoV-2 genome
sequencing dataset to date --- the largest of any virus in history,
with millions of sequences.  This unprecedented amount of genomic data
and easy availability allowed researchers to explore the
molecular mechanism, genetic variability, evolutionary progress, and
capability of development and spread of novel variants of the virus.
On the other hand, this amount of data exceeds the capacity of the  more traditional phylogenetic methods
such as Nextstrain~\citep{hadfield2018a} or even the more recent
IQTREE2~\citep{minh_2020_iqtree2} by several orders of magnitude --- a
Big Data challenge.  As a result, recent alternative approaches based
on clustering and classification of sequences, \eg, to identify major
variants have appeared in the
literature~\citep{kuzmin2020machine,melnyk2020clustering1,ali2021k,ali2021effective,ali2021spike2vec},
with promising accuracy and scalability properties.

Many issues still remain, however, such as sequencing errors being
mistaken for mutations in different analyses when studying the
evolutionary and transmission patterns of
SARS-CoV-2~\citep{gisaid_history,arons2020presymptomatic}, or other
viruses.  Incorporation of error in NGS sequences due to contamination
in sample preparation, sequencing technology, or genome assembly
methodology are other confounding factors.  Generally, computational
biologists filter those sequences having errors or mask those sequence
fragments having errors.  For example, each
GISAID~\citep{gisaid_website_url} sequence is a consensus sequence
from the intra-host viral population sampled from the patient,
averaging out the minor variations which exist in this population.
While such a consensus sequence is a good representative of this
population, \ie, it is still precise enough to capture the SARS-CoV-2
variant harbored by the infected individual, it comes at the cost of
losing this important information, such as these minor variations.
Such minor variations, when given enough time to evolve, \eg, within
an immunocompromised individual can become dominant --- one of the
theories behind the emergence of the Alpha
variant~\citep{frampton2021genomic}.

Many machine learning approaches towards clustering and classification of
sequences~\citep{kuzmin2020machine,ali2021k,ali2021effective,ali2021spike2vec} have been
operating under somewhat idealized conditions of virtually error-free
consensus sequences, which may not be in certain settings.  Moreover, some of these methods rely on a $k$-mer based
feature vector representation --- an approach that does not even rely
on alignment of the sequences, which may not always be available in certain settings, and can also introduce
bias~\citep{golubchik2007mind}.  Such a framework should easily cope
with errors as well --- something machine learning approaches can do
very naturally~\citep{du2021improving}.  Hence, there is a great need
for some way to reliably benchmark such methods for robustness to
errors, which is what we carry out in this paper. Our main research question is the following:

\begin{center}
   \textit{How robust are the existing classification models to adversarial coronavirus inputs?} 
\end{center}

In this paper, we extend our error testing procedure as a framework for
benchmarking the performance of different ML methods in terms of
classification accuracy and robustness to different types of simulated nucleotide sequences with errors.  This involves using PBSIM and InSilicoSeq tools for simulating long reads and short reads based nucleotide sequences with realistic error profiles.

We highlight the main contributions of this paper as follows: 
\begin{itemize}
  \item We propose several ways of introducing biologically meaningful errors into the SARS-CoV-2 genome sequences, which reflect the error profiles of modern NGS technologies such as Illumina and PacBio.
  \item We propose two different embedding methods, called PSSM2Vec (based on the concept of position weight matrix) and Min2Vec (based on the concept of minimizers), to generate a fixed length numerical representation of biological sequences. Using these embedding methods, we perform classification on original and errored sequences and report the performance using different evaluation metrics.
  \item We show that the alignment free method for feature embedding, called Spike2Vec, which is based on the idea of $k$-mers is better in terms of predictive performance when there is no error in the nucleotide sequences.  This is likely due to the fact that it preserves nucleotide order information in more detail than the PSSM2Vec or Min2Vec representations (at the expense of being less compact).
  \item We demonstrate that for the PBSIM (long reads) based errored sequences, that the PSSM2Vec embedding is more robust than the sliding window based Spike2Vec or Min2Vec approaches, possibly because it captures more long-range information.
  \item We show that for the Illumina based errored sequences, Spike2Vec and Min2Vec are able to show better performance than PSSM2Vec, again because they likely preserve order information in higher detail.
\end{itemize}

This rest of the paper is organized as follows.  In Section~\ref{sec:related} we
discuss related work.  The methods to generate the adversarial examples is described in Section~\ref{sec:adversarial}.
In Section~\ref{sec:approach}, we discuss different embedding methods used to convert the sequences into fixed-length numerical representation.  Section~\ref{sec:experiments} contains the details regarding the
experimental setup, dataset statistics, and data visualization. We propose our results for accuracy and robustness in Section.~\ref{sec:results}.  We described the limitations of our work in Section~\ref{sec_limitations}.
Finally, we conclude this paper in Section~\ref{sec:conclusion}.

\section{Related Work}
\label{sec:related}

Assessing and benchmarking the robustness of ML or DL approaches by a
series of adversarial attacks are popular in the image classification
domain~\citep{hendrycks2019benchmarking}, but there are others that
are closer to the domain of molecular data.
In~\citep{schwalbe2021differentiable}, the authors provide a series of
realistic adversarial attacks to benchmark methods that predict
chemical properties from atomistic simulations \eg, molecular
conformation, reactions, and phase transitions. Even closer to the
subject of our paper --- protein sequences -- the authors
of~\citep{jha2021robust} show that methods, such as
AlphaFold~\citep{jumper2021alpha} and
RoseTTAFold~\citep{baek2021rosetta} which employ deep neural networks
to predict protein conformation are not robust: producing drastically
different protein structures as a result of very small biologically
meaningful perturbations in the protein sequence. Our approach is
similar, albeit with a different goal of classification: namely, to
explore how a small number of mutations (simulating the error
introduced certain types of NGS technologies) can affect the
downstream classification of different machine learning and deep
learning approaches. 
After getting the numerical representation, a popular approach is to get the kernel matrix and give that matrix as input to traditional machine learning classifiers like support vector machines (SVM)~\citep{leslie2002mismatch,farhan2017efficient,Kuksa_SequenceKernel}. However, these methods are expensive in terms of space complexity. Authors in~\citep{ali2021spike2vec,kuzmin2020machine} propose an efficient embedding method for classification and clustering of spike sequences. However, their approaches are either not scalable or perform poorly on bigger datasets.

\section{Adversarial Sequence Creation}
\label{sec:adversarial}

We use two types of approaches to generate adversarial examples so that we can test the robustness of different machine learning methods. 

\subsection{PBSIM simulated data generation}
PBSIM is developed to simulate Pacific Biosciences (PacBio)  sequencing reads. Generally, the PacBio sequencer generates two types of reads: continuous long read (CLR) and circular consensus sequencing short reads (CCS). The CLR reads have a high error rate, and CCS reads have a lower error rate. 
PBSIM can simulate both CLR and CCS reads with different approaches: sampling-based simulation and model-based simulation. In the sampling-based simulation, PBSIM considers the length and quality of a provided read set to simulate the reads. In model-based simulation, PIBSIM simulates the reads on the basis of an error model \citep{ono_2012_pbsim}.

To generate a sequence with errors, we take an original SARS-CoV-2 genomic sequence and simulate reads from it using the model-based approach, with the default PacBio error model.
These reads (containing errors) are then aligned to the original sequence, mutations are called, and then consensus sequences (with mutations, some of which are errors) are extracted.
We control the amount of error in the reads by adjusting the depth of the reads (a parameter of PBSIM).
We generate such a sequence for 8220 different SARS-CoV-2 sequences from GISAID, for read depths 5 and 10 (more depths appear in the supplement).

\subsection{InSilicoSeq simulated data generation}
The InSilicoSeq open-source tool simulates the reads from different short read technologies such as Illumina.
InSilicoSeq is a widely used tool, and several studies generate more realistic NGS data using this tool for planning new experiments and benchmarking purposes \citep{kalantar_2020_idseqan,sangiovanni_2019_from,whibley_2021_the,andreusnchez_2021_a,glickman_2021_simulation}.
The tool can incorporate errors into the reads based on the details (e.g., chemistry) of recent Illumina platforms.
InSilicoSeq supports substitution, insertion, and deletion errors and can model PHRED score.
The current release of the InSilicoSeq tool has a pre-built error model for HiSeq, MiSeq, and NovaSeq instruments.
Moreover, InSilicoSeq has the option to generate the number of reads according to the user's requirement \citep{gourl_2018_simulating}. 

We generate a sequence with errors analogously to the above, this time controlling the error by adjusting this number of reads.  We generate a sequence for the 8220 GISAID sequences mentioned above, with a number of reads 5000 and 10000 (with more numbers of reads in the supplement).

\section{Feature Embeddings Generation}
\label{sec:approach}



In this section, we introduce different feature embedding methods used to convert the nucleotide sequence into a fixed length representation.


\subsection{Spike2Vec~\citep{ali2021spike2vec}}
\begin{figure}[h!]
    \centering
    \includegraphics[scale = 0.3] {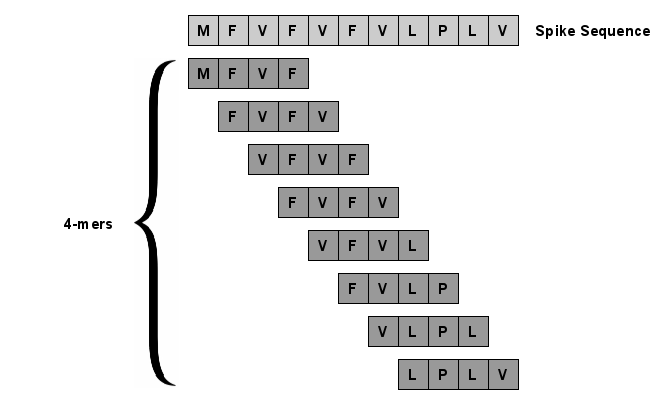}
    \caption{Example of different length $k$-mers in a protein (spike) sequence ``MDPEG"~\citep{ali2021spike2vec}}
    \label{fig_k_mers}
\end{figure}

A popular approach to preserve the ordering of the sequential information, called Spike2Vec~\citep{ali2021spike2vec}, takes the sliding window based substrings (called mers) of length $k$ (also called $n$-gram). 
This $k$-mers based representation helps to preserve the order of characters within the sequences~\citep{ali2021k} (see Figure~\ref{fig_k_mers} for example of $k$-mers).
In this approach, first, the $k$-mers are computed for each nucleotide sequence. Then a fixed length frequency vector is generated corresponding to each sequence, which contains the count of each $k$-mer in that sequence. One advantage of using $k$-mers based approach is that it is an \textit{``alignment free"} method unlike other popular baselines (e.g. One-Hot Encoding ``OHE"~\citep{kuzmin2020machine,ali2021k}), which requires the sequences to be aligned.
Note that sequence alignment is an expensive process and requires a reference
sequence (genome)~\citep{chowdhury2017review,denti2021shark}.  It may
also introduce bias into the result~\citep{golubchik2007mind}.
The total number of $k$-mers in a given nucleotide sequences are
    $N - k + 1$,
where $N$ is the length of the sequence. The variable $k$ is the user defined parameter. In this paper, we take $k = 3$ (decided empirically using standard validation set approach~\citep{validationSetApproach}).

\subsubsection{Frequency Vector Generation}\label{sec_freq_vec_gen}
After generating the $k$-mers, the next step is to generate the fixed-length numerical representation (frequency vector) for the set of $k$-mers in a nucleotide sequence. Suppose the set of nucleotides in the whole dataset is represented by alphabet $\Sigma$ (A, C, G, and T). Now, the length of the frequency vector  will be $\vert \Sigma \vert^k$ (all possible combinations of $k$-mers in $\Sigma$ of length $k$). 
In our dataset, since we have $4$ unique nucleotides in any sequence, the length of frequency vector in our case would be $4^3 = 64$ (when we take $k = 3$).

\subsection{PSSM2Vec}
The PSSM2Vec embedding is based on the idea of Position Specific Scoring Matrix (PSSM), also called position weight matrix (PWM)~\citep{stormo1982pwm,ali2022pwm2vec}. For a given nucleotide sequence $s$, PSSM2Vec designs the PWM. The PWM generation start by first computing the $k$-mers (where $k = 3$, which is decided using standard validation set approach~\citep{validationSetApproach}) for $s$. For all the $k$-mers in $s$, a matrix of length $\vert \Sigma \vert \times k$ is generated, which includes the count of nucleotides at different positions within the $k$-mers. This matrix is also called the Position Frequency Matrix (PFM). In the next step, column wise probabilities are computed for PFM to get a new matrix, called the Position Probability Matrix (PPM). More formally, PPM is computed as follows:
    $\frac{\text{Frequency of nucleotide}}{\text{No. of
        nucleotides in the column}}$.
To avoid having zero in the denominator, we add a small value of $0.01$ (called Laplace value or pseudocount) during probability computation. Finally, the PWM is computed from the PPM by taking the log likelihood of each nucleotide $c \in \Sigma$ at a position $i$. More formally:
\begin{equation}
    W_{c, i}=\log_{2} \frac{p(c, i)}{p(c)}
\end{equation}
where $p(c) = \frac{1}{4}$, which corresponds to the equal probability of occurrence for each nucleotide in the sequence. After generating the PWM, we flatten the matrix to generate a single vector, which we refer to as PSSM2Vec.

\subsection{Min2Vec}
The Min2Vec feature embedding is based on the idea of minimizers~\citep{roberts-2004-minimizer}. The minimizers is a modified version of $k$-mers and used to represent the biological sequences in more compact form.
\begin{definition}[Minimizers]
For a given $k$-mer, a minimizer (also called m-mer) is a substring of consecutive nucleotides of length $m$ from the $k$-mer, which is lexicographically smallest one in both forward and backward order of the k-mer, where $m<k$ and is fixed.
\end{definition}
The pseudocode to compute the minimizers is given in Algorithm~\ref{algo_minimizer}. For better understanding of pseudocode, we use the syntax of python code. To compute the minimizer, we take $k=9$ and $m=3$, which is decided empirically using standard validation set approach~\citep{validationSetApproach}.
After computing the minimizers for a given nucleotide sequence, we follow the same method to generate the frequency vector based representation as describe in Section~\ref{sec_freq_vec_gen}. For reference, we call this method as Min2Vec.

\begin{algorithm}[h!]
	\caption{Minimizer Computation}
\label{algo_minimizer}
	\begin{algorithmic}[1]
	\State \textbf{Input:} Sequence $s$ and integer $k$ and $m$
	\State \textbf{Output:} Set of Minimizers
	
	\State minimizers = $\emptyset$
    \State queue = [] \Comment{$ \text{maintain queue of all m-mers }$}
    \State idx = 0  \Comment{ $ \text{index of the current minimizer}$}
     \For{$i \leftarrow 1 \textup { to }|s|-k+1$}   
        \State kmer = $s[i:i+k]$ 

        \If{\textup{idx} $> 1$} 
            \State queue.dequeue 
            \State mmer = $s[i+k-m:i+k]$  \Comment{$  \text{new m-mer}$}
            \State idx $\leftarrow$ idx $- 1$  \Comment{$  \text{shift index of current minimizer}$}
            \State mmer = min(mmer, reverse(mmer))  \Comment{$  \text{lexicographically smallest forw./rever.}$}
            \State queue.enqueue(mmer) 
            
            \If{\textup{mmer $<$ queue[idx]}}
                \State idx = $k-m$ \Comment{$  \text{update minimizer with new m-mer}$}
             \EndIf
        \Else
                
                \State queue = [] \Comment{$  \text{reset the queue}$}
                \State idx = 0
                \For{$j \leftarrow 1 \textup{ to } k-m+1$}
                    \State mmer = kmer$[j:j+m]$ \Comment{$ \text{compute each m-mer}$}
                    \State mmer = min(mmer, reverse(mmer))
                    \State queue.enqueue(mmer)
                    \If{\textup{mmer $<$ queue[idx]}}
                        \State idx = $j$ \Comment{$ \text{index of current minimizer}$}
                    \EndIf
                \EndFor
        \EndIf
        
        \State minimizers $\leftarrow$ minimizers $\cup$ queue[idx] \Comment{$ \text{add current minimizer}$}
    \EndFor 
        \State return(minimizers)
	\end{algorithmic}
\end{algorithm}

\section{Experimental Setup}
\label{sec:experiments}

All experiments are conducted using an Intel(R) Core i5 (11th generation) with $2.40$ GHz processor having windows $10$ ($64$ bit) OS and $32$ GB memory.
The simulated and pre-processed data is available online~\footnote{\url{https://github.com/sarwanpasha/Adversarial_attack_on_biological_sequences}}.
For classification purpose, we use Support Vector Machine (SVM), Naive Bayes (NB), Multi-Layer Perceptron (MLP), K-Nearest Neighbors (KNN), Random Forest (RF), Logistic Regression (LR), and Decision Tree (DT). 

To measure the performance of ML models, we apply two different strategies. The first strategy is called Accuracy. In this case, we compute the average accuracy, precision, recall, F1 (weighted), F1 (Macro), and ROC-AUC for the whole (original) dataset without any errored sequence. The second strategy is called robustness. In this case, we only consider the adversarial examples (set of errored sequences) for the test set (and non errored sequences  for training set) and compute average accuracy, precision, recall, F1 (weighted), F1 (Macro), and ROC-AUC for the ML models.

\subsection{Dataset Statistics}\label{sub_sec_dataset_stats}
We used the full length nucleotide sequences of coronavirus from a popular and publicly available database of SARS-CoV-2, GISAID~\footnote{\url{https://www.gisaid.org/}}.  In our dataset, we have $8220$ nucleotide sequences along with the COVID-19 variant information. The total number of unique variants in our dataset are $41$.
The dataset statistics for the prepossessed data are given in
Table~\ref{tbl_variant_information} (in appendix). Our simulated dataset is available  online~\footnote{\url{https://drive.google.com/drive/folders/1adtr8FImIYTqxM20wgInRqIZ8EJY4HVS?usp=sharing}} for reproducibility. We also provide the visualization of different embedding methods (using t-SNE~\citep{van2008visualizing}) in Appendix~\ref{sec_data_viz}.

\section{Results and Discussion}
\label{sec:results}
In this section, we show the accuracy results for the original data followed by the robustness results of PBSIM and Illumina based errored sequences. 

\subsection{Accuracy Results}
To evaluate the performance of original nucleotide sequences (non-errored sequences), we split the original $8220$ sequences into (random) $70-30 \%$ training and testing set and perform classification on different embedding methods. We ran the experiments 5 times and report average results.
The accuracy results for different embeddings and ML models are shown in Table~\ref{tbl_acc_org}. We can observe that the SVM classifiers with Spike2Vec based embedding outperform other embeddings and ML models for all but one evaluation metrics. In terms of runtime, since the length of vectors for PSSM2Vec is smaller than the other embedding methods, its training runtime for NB classifier is the smallest one.
\begin{table}[h!]
  \centering
  \caption{Accuracy Results on 8220 (original) nucleotide sequences (without any error). Best values are shown in bold.}
    \resizebox{0.7\textwidth}{!}{
  \begin{tabular}{p{1.9cm}p{0.8cm}p{0.8cm}p{0.8cm}p{0.7cm}p{0.7cm}cp{0.7cm} | p{1.1cm}}
    \hline
    \multirow{3}{1.1cm}{Embed. Method} & \multirow{3}{0.7cm}{ML Algo.} & \multirow{3}{*}{Acc.} & \multirow{3}{*}{Prec.} & \multirow{3}{*}{Recall} & \multirow{3}{0.9cm}{F1 weigh.} & \multirow{3}{0.9cm}{F1 Macro} & \multirow{3}{1.2cm}{ROC- AUC} & Train. runtime (sec.) \\	
    \hline	\hline	

    \multirow{7}{*}{Spike2Vec}  
     & SVM & \textbf{0.87} & \textbf{0.87} & \textbf{0.87} & \textbf{0.86} & \textbf{0.76} & \textbf{0.87} & 7.43 \\
     & NB & 0.03 & 0.05 & 0.03 & 0.02 & 0.05 & 0.55 & 0.09 \\
     & MLP & 0.75 & 0.74 & 0.75 & 0.74 & 0.36 & 0.68 & 18.42 \\
     & KNN & 0.73 & 0.73 & 0.73 & 0.71 & 0.48 & 0.71 & 2.04 \\
     & RF & 0.82 & 0.85 & 0.82 & 0.80 & 0.67 & 0.78 & 2.17 \\
     & LR & 0.86 & 0.85 & 0.86 & 0.85 & 0.70 & 0.84 & 8.67 \\
     & DT & 0.67 & 0.67 & 0.67 & 0.66 & 0.42 & 0.71 & 0.27 \\

    \cline{2-9}	
    \multirow{7}{*}{PSSM2Vec}  
    & SVM & 0.28 & 0.08 & 0.28 & 0.12 & 0.01 & 0.50 & 3.14 \\
    & NB & 0.01 & 0.01 & 0.01 & 0.00 & 0.01 & 0.52 & \textbf{0.03} \\
    & MLP & 0.34 & 0.27 & 0.34 & 0.26 & 0.06 & 0.53 & 17.31 \\
    & KNN & 0.32 & 0.28 & 0.32 & 0.28 & 0.13 & 0.55 & 0.33 \\
    & RF & 0.33 & 0.30 & 0.33 & 0.31 & 0.16 & 0.57 & 1.60 \\
    & LR & 0.28 & 0.08 & 0.28 & 0.12 & 0.01 & 0.50 & 0.68 \\
    & DT & 0.29 & 0.28 & 0.29 & 0.28 & 0.13 & 0.56 & 0.06 \\
    \cline{2-9}	
    \multirow{7}{*}{Min2Vec}  
     & SVM & 0.60 & 0.58 & 0.60 & 0.56 & 0.48 & 0.72 & 15.19 \\
     & NB & 0.05 & 0.12 & 0.05 & 0.04 & 0.12 & 0.59 & 0.08 \\
     & MLP & 0.57 & 0.52 & 0.57 & 0.53 & 0.30 & 0.64 & 26.32 \\
     & KNN & 0.55 & 0.56 & 0.55 & 0.53 & 0.37 & 0.66 & 1.51 \\
     & RF & 0.75 & 0.79 & 0.75 & 0.74 & 0.61 & 0.76 & 1.72 \\
     & LR & 0.58 & 0.55 & 0.58 & 0.54 & 0.40 & 0.68 & 6.36 \\
     & DT & 0.64 & 0.64 & 0.64 & 0.64 & 0.48 & 0.74 & 0.14 \\

    \hline
  \end{tabular}
  }
  \label{tbl_acc_org}
\end{table}

\subsection{Robustness Results}
For the robustness results, we show the predictive performance of different ML models by first using the PBSIM based adversarial sequences and then show the performance of ML models for Illumina based adversarial examples.

For the PBSIM based sequences, we take the original $8220$ (non-errored) sequence data for training the ML models and use the PBSIM based ($8220$) errored sequences as the test set. The purpose of this experimental setting is to evaluate the performance of ML models on the errored sequences, which were not available during the training process.
In this experimental setting, we show the results for depth $5$ and depth $10$ based errored sequences (in test set) in Table~\ref{tbl_pbsim_depth_5}.

\begin{table}[h!]
  \centering
  \caption{Robustness Results on PBSIM data with $5$ and $10$ as read depth.}
    \resizebox{0.95\textwidth}{!}{
  \begin{tabular}{@{\extracolsep{4pt}}p{1.9cm}p{0.5cm}p{0.4cm}p{0.4cm}p{0.7cm}p{0.7cm}cp{0.7cm} p{1.1cm} p{0.4cm}p{0.4cm}p{0.7cm}p{0.7cm}cp{0.7cm} p{1.1cm}}
    \hline
    & & \multicolumn{7}{c}{Depth: 5} & \multicolumn{7}{c}{Depth: 10}
    \\
    \cline{3-9} \cline{10-16}
    \multirow{3}{1.1cm}{Embed. Method} & \multirow{3}{0.7cm}{ML Algo.} & \multirow{3}{*}{Acc.} & \multirow{3}{*}{Prec.} & \multirow{3}{*}{Recall} & \multirow{3}{0.9cm}{F1 weigh.} & \multirow{3}{0.9cm}{F1 Macro} & \multirow{3}{1.2cm}{ROC- AUC} & Train. runtime (sec.) & \multirow{3}{*}{Acc.} & \multirow{3}{*}{Prec.} & \multirow{3}{*}{Recall} & \multirow{3}{0.9cm}{F1 weigh.} & \multirow{3}{0.9cm}{F1 Macro} & \multirow{3}{1.2cm}{ROC- AUC} & Train. runtime (sec.) \\	
    \hline	\hline	
    \multirow{7}{*}{Spike2Vec}  
     & SVM & 0.01 & 0.00 & 0.01 & 0.00 & 0.00 & 0.50 & 16.48  & 0.01 & 0.00 & 0.01 & 0.00 & 0.00 & 0.50 & 16.88 \\
    & NB & 0.00 & 0.00 & 0.00 & 0.00 & 0.00 & 0.50 & 0.68 & 0.00 & 0.00 & 0.00 & 0.00 & 0.00 & 0.50 & 0.71 \\
    & MLP & 0.28 & 0.08 & 0.28 & 0.12 & 0.01 & 0.50 & 23.65 & 0.02 & 0.00 & 0.02 & 0.00 & 0.00 & 0.50 & 16.86 \\
    & KNN & 0.28 & 0.08 & 0.28 & 0.12 & 0.01 & 0.50 & 1.68 & 0.28 & 0.08 & 0.28 & 0.12 & 0.01 & 0.50 & 1.78 \\
    & RF & 0.28 & 0.08 & 0.28 & 0.12 & 0.01 & 0.50 & 1.78 & 0.28 & 0.08 & 0.28 & 0.12 & 0.01 & 0.50 & 2.88 \\
    & LR & 0.01 & 0.00 & 0.01 & 0.00 & 0.00 & 0.50 & 11.30 & 0.01 & 0.00 & 0.01 & 0.00 & 0.00 & 0.50 & 12.04 \\
    & DT & 0.01 & 0.00 & 0.01 & 0.00 & 0.00 & 0.50 & 0.34  & 0.01 & 0.00 & 0.01 & 0.00 & 0.00 & 0.50 & 0.36 \\
    \cline{2-9}	\cline{10-16}	
    \multirow{7}{*}{PSSM2Vec}  
    & SVM & 0.27 & 0.07 & 0.27 & 0.11 & 0.01 & 0.50 & 8.14 & 0.30 & 0.09 & 0.30 & 0.13 & 0.01 & 0.50 & 8.32 \\
 & NB & 0.27 & 0.07 & 0.27 & 0.11 & 0.01 & 0.50 & 0.34 & 0.30 & 0.09 & 0.30 & 0.13 & 0.01 & 0.50 & 0.36 \\
 & MLP & 0.27 & 0.07 & 0.27 & 0.11 & 0.01 & 0.50 & 7.47 & 0.30 & 0.09 & 0.30 & 0.13 & 0.01 & 0.50 & 7.90 \\
 & KNN & 0.27 & 0.07 & 0.27 & 0.11 & 0.01 & 0.50 & 0.51 & 0.01 & 0.05 & 0.01 & 0.00 & 0.00 & 0.50 & 0.52 \\
 & RF & 0.27 & 0.07 & 0.27 & 0.11 & 0.01 & 0.50 & 1.17 & 0.30 & 0.09 & 0.30 & 0.13 & 0.01 & 0.50 & 0.98 \\
 & LR & 0.27 & 0.07 & 0.27 & 0.11 & 0.01 & 0.50 & 3.76 & 0.30 & 0.09 & 0.30 & 0.13 & 0.01 & 0.50 & 3.62 \\
 & DT & 0.27 & 0.07 & 0.27 & 0.11 & 0.01 & 0.50 & 0.02 & 0.30 & 0.09 & 0.30 & 0.13 & 0.01 & 0.50 & 0.02 \\
    \cline{2-9}	\cline{10-16}	
    \multirow{7}{*}{Min2Vec}  
    & SVM & 0.27 & 0.07 & 0.26 & 0.11 & 0.01 & 0.50  &  5.22 & 0.27 & 0.08 & 0.27 & 0.12 & 0.01 & 0.50 & 4.91 \\
    & NB & 0.26 & 0.07 & 0.27  & 0.11 & 0.26 & 0.50 & 0.43  & 0.27 & 0.08 & 0.27 & 0.12 & 0.01 & 0.50 & 0.34 \\
 &  MLP & 0.26 & 0.07 & 0.26 & 0.11 & 0.26 & 0.50 & 1.63  & 0.27 & 0.08 & 0.27 & 0.12 & 0.01 & 0.50 & 1.92 \\
 &  KNN & 0.26 & 0.07 & 0.26 & 0.11 & 0.26 & 0.50 & 0.62 & 0.08 & 0.01 & 0.08 & 0.01 & 0.00 & 0.50 & 0.69 \\
 &  RF & 0.26 & 0.07 & 0.26 & 0.11 & 0.26 & 0.50 & 0.67 & 0.27 & 0.08 & 0.27 & 0.12 & 0.01 & 0.50 & 0.77 \\
 &  LR & 0.26 & 0.07 & 0.26 & 0.11 & 0.26 & 0.50 & 0.69 & 0.27 & 0.08 & 0.27 & 0.12 & 0.01 & 0.50 & 0.67 \\
 &  DT & 0.26 & 0.07 & 0.26 & 0.11 & 0.26 & 0.50 & 0.17 & 0.27 & 0.08 & 0.27 & 0.12 & 0.01 & 0.50 & 0.26 \\
    \hline
  \end{tabular}
  }
  \label{tbl_pbsim_depth_5}
\end{table}

For the Illumina based sequences, we take the original $8220$ (non-errored) sequence data for training the ML models and use the illumina based errored $8220$ sequences as the test set. In this experimental setting, we show the results for sequences simulated using different number of short reads (in test set). The results for $5000$ short reads and $10000$ short reads based errored sequences are shown in Table~\ref{tbl_illumina_5k_reads}.

\begin{table}[h!]
  \centering
   \caption{Robustness Results on Illumina based errored sequences with $5000$ and $10000$ short reads used in the simulation process.}
    \resizebox{0.95\textwidth}{!}{
  \begin{tabular}{@{\extracolsep{4pt}}p{1.9cm}p{0.5cm}p{0.4cm}p{0.4cm}p{0.7cm}p{0.7cm}cp{0.7cm} p{1.1cm} p{0.4cm}p{0.4cm}p{0.7cm}p{0.7cm}cp{0.7cm} p{1.1cm}}
    \hline
    & & \multicolumn{7}{c}{\# of Short Reads: 5000} & \multicolumn{7}{c}{\# of Short Reads: 10000}
    \\
    \cline{3-9} \cline{10-16}
    \multirow{3}{1.1cm}{Embed. Method} & \multirow{3}{0.7cm}{ML Algo.} & \multirow{3}{*}{Acc.} & \multirow{3}{*}{Prec.} & \multirow{3}{*}{Recall} & \multirow{3}{0.9cm}{F1 weigh.} & \multirow{3}{0.9cm}{F1 Macro} & \multirow{3}{1.2cm}{ROC- AUC} & Train. runtime (sec.) & \multirow{3}{*}{Acc.} & \multirow{3}{*}{Prec.} & \multirow{3}{*}{Recall} & \multirow{3}{0.9cm}{F1 weigh.} & \multirow{3}{0.9cm}{F1 Macro} & \multirow{3}{1.2cm}{ROC- AUC} & Train. runtime (sec.) \\	
    \hline	\hline	
    \multirow{7}{*}{Spike2Vec}  
     & SVM & 0.68 & 0.66 & 0.68 & 0.66 & 0.49 & 0.73 & 6.75 & 0.73 & 0.72 & 0.71 & 0.72 & 0.55 & 0.76 & 10.76 \\
 &  NB & 0.69 & 0.73 & 0.69 & 0.71 & 0.57 & 0.80 & 0.31 & 0.72 & 0.72 & 0.72 & 0.72 & 0.53 & 0.77 & 0.32 \\
 &  MLP & 0.68 & 0.65 & 0.68 & 0.66 & 0.34 & 0.66 & 75.93 & 0.68 & 0.65 & 0.68 & 0.66 & 0.32 & 0.65 & 27.84 \\
 &  KNN & 0.73 & 0.73 & 0.73 & 0.72 & 0.57 & 0.76 & 0.75 & 0.73 & 0.72 & 0.73 & 0.72 & 0.56 & 0.76 & 0.68 \\
 &  RF & 0.72 & 0.72 & 0.72 & 0.70 & 0.51 & 0.72 & 2.44 & 0.73 & 0.73 & 0.73 & 0.71 & 0.55 & 0.74 & 2.43 \\
 &  LR & 0.72 & 0.70 & 0.72 & 0.70 & 0.52 & 0.74 & 6.71 & 0.72 & 0.71 & 0.72 & 0.71 & 0.54 & 0.75 & 6.69 \\
 &  DT & 0.51 & 0.53 & 0.51 & 0.52 & 0.32 & 0.66 & 0.24 & 0.56 & 0.56 & 0.56 & 0.56 & 0.41 & 0.70 & 0.21 \\
    \cline{2-9}	\cline{10-16}	
    \multirow{7}{*}{PSSM2Vec}  
    & SVM & 0.27 & 0.07 & 0.27 & 0.12 & 0.01 & 0.50 & 8.20 & 0.28 & 0.08 & 0.28 & 0.12 & 0.01 & 0.50  &  9.64 \\
 &  NB & 0.01 & 0.00 & 0.01 & 0.00 & 0.01 & 0.51 & 0.39 & 0.02 & 0.01 & 0.02 & 0.01 & 0.03 & 0.52 & 0.25 \\
 &  MLP & 0.32 & 0.22 & 0.32 & 0.24 & 0.06 & 0.52 & 10.30 & 0.34 & 0.25 & 0.34 & 0.26 & 0.08 & 0.53 & 12.72 \\
 &  KNN & 0.26 & 0.21 & 0.26 & 0.22 & 0.06 & 0.52 & 1.10 & 0.29 & 0.26 & 0.29 & 0.25 & 0.09 & 0.54 & 0.70 \\
 &  RF & 0.30 & 0.24 & 0.30 & 0.25 & 0.08 & 0.52 & 2.17 & 0.32 & 0.25 & 0.32 & 0.27 & 0.08 & 0.53 & 1.92 \\
 &  LR & 0.27 & 0.07 & 0.27 & 0.12 & 0.01 & 0.50 & 3.92 & 0.28 & 0.08 & 0.28 & 0.12 & 0.01 & 0.50 & 3.26 \\
 &  DT & 0.30 & 0.24 & 0.30 & 0.25 & 0.07 & 0.52 & 0.12 & 0.32 & 0.25 & 0.32 & 0.26 & 0.08 & 0.53 & 0.07 \\
    \cline{2-9}	\cline{10-16}	
    \multirow{7}{*}{Min2Vec}  
   & SVM & 0.52 & 0.47 & 0.52 & 0.46 & 0.30 & 0.64 & 11.75 & 0.54 & 0.50 & 0.54 & 0.49 & 0.34 & 0.66 & 7.45 \\
    & NB & 0.05 & 0.27 & 0.05 & 0.04 & 0.09 & 0.63 & 0.20 & 0.07 & 0.37 & 0.07 & 0.08 & 0.14 & 0.64 & 0.19 \\
 &  MLP & 0.52 & 0.46 & 0.52 & 0.46 & 0.26 & 0.62 & 25.0  & 0.52 & 0.46 & 0.52 & 0.48 & 0.25 & 0.62 & 28.70 \\
 &  KNN & 0.55 & 0.55 & 0.55 & 0.53 & 0.39 & 0.67 & 0.52 & 0.57 & 0.57 & 0.57 & 0.56 & 0.47 & 0.70 & 0.56 \\
 &  RF & 0.65 & 0.67 & 0.65 & 0.63 & 0.46 & 0.70 & 1.75 & 0.68 & 0.69 & 0.68 & 0.66 & 0.56 & 0.74 & 1.60 \\
 &  LR & 0.51 & 0.46 & 0.51 & 0.46 & 0.28 & 0.63 & 2.91 & 0.53 & 0.49 & 0.53 & 0.48 & 0.34 & 0.65 & 2.90 \\
 &  DT & 0.47 & 0.47 & 0.47 & 0.47 & 0.31 & 0.65 & 0.12  & 0.54 & 0.54 & 0.54 & 0.54 & 0.42 & 0.70 & 0.10 \\
    \hline
  \end{tabular}
  }
  \label{tbl_illumina_5k_reads}
\end{table}

\subsection{PBSIM vs Illumina Results Discussion}
Third generation sequencing technologies such as PacBio and Oxford Nanopore Technologies (ONT), being newer than traditional high-throughput NGS technologies (e.g., Illumina), offer longer reads, which are useful to efforts such as haplotype assembly~\citep{patterson-2015-whatshap,beretta-2018-hapchat}.  The drawback with these technologies is that they have lower coverage and contain more errors --- up to $15\%$ error rate as compared to the less than $1\%$ with state-of-the-art Illiumina~\citep{weirather_2017_comprehensive,fu_2019_a,stoler_2021_sequencing,ma_2019_analysis}. Therefore, it is not surprising that perturbing the coverage in the case of Pacbio (PBSIM) based experiment had a larger effect (see Table~\ref{tbl_pbsim_depth_5}) on the predictive performance of ML models as compared to the Illumina (insilicoseq) based experiment (see Table~\ref{tbl_illumina_5k_reads}).  The sequences submitted to GISAID~\footnote{\url{https://www.gisaid.org/}} database are almost exclusively from high-throughput technologies~\citep{gisaid_website_url}, hence we got more stable results on the original sequences (without adding any additional error) extracted from GISAID (see Table~\ref{tbl_acc_org}). 

For the PBSIM based errored sequences, we can observe that PSSM2Vec outperforms the other two embedding methods (see Table~\ref{tbl_pbsim_depth_5}), which means that a sliding window based approach (using $k$-mers or $m$-mers) is not desirable while dealing with Pacbio errors.
This could be due to the fact that the PSSM2Vec representation captures more long-range information than the shorter (length $k$) sliding window.
Similarly, for the Illumina based sequences, we can observe the opposite behavior (see Table~\ref{tbl_illumina_5k_reads}), where the sliding window based approaches are better than the position weight matrix based embedding.
This could be because in PSSM2Vec, the order of nucleotides is not preserved in as much detail (because we just take the position weight matrix and make it a 1-D vector by flattening it).
In the sliding window based approach, the majority of the nucleotides appear in $k$ sliding windows, hence preserving more (order) information, which results in better predictive performance (because of less loss of information in generating the numerical embedding).
This comes at the cost of it being a higher dimensional representation, of course.

\section{Limitations}\label{sec_limitations}
We used feature engineering based embeddings for the experiments in this paper. However, using neural network could improve the accuracy and/or robustness of the benchmark dataset. Moreover, we use the Illumina based data with $5000$ and $10000$ short reads only --- we believe that using a larger number of reads may improve the performance of the underlying classifier.  The same is true for PBSIM data where we use only 5 and 10 as read depth. Please note that results on more number of short reads (for illumina) and read depth (for PBSIM) are given in the supplementary material file.

\section{Conclusion}
\label{sec:conclusion}
In this paper, we propose two different ways to test the robustness of ML models in terms of sequence classification. We test the accuracy and robustness of ML models using different embedding methods.
We conclude that for different simulation tools, different embedding methods perform better than the others and there is no clear winner, which consistently outperforms in all scenarios.
One interesting future extension is to use other embedding methods from the literature and also apply deep learning models for classification of sequences. Studying adversarial attack on other viruses (e.g. Zika) is also an interesting future extension.


\bibliographystyle{IEEEtran}
\bibliography{references}


\pagebreak

\appendix

\section{Appendix A: Dataset Statistics}
The information regarding the number of variants (used as class labels) in our data are given in Table~\ref{tbl_variant_information}.
\begin{table}[ht!]
  \centering
  \caption{Dataset Statistics for different variants in our data. The total number of sequences (and corresponding variants) are $8220$ after preprocessing.}
  \begin{tabular}{cc|cc}
    \hline
    
      \multirow{1}{1.2cm}{Variant} & \multirow{1}{*}{Num. of sequences} & \multirow{1}{1.2cm}{Variant} & \multirow{1}{*}{Num. of sequences}\\
      \hline	\hline
        AY.103   & 2271   & AY.121 & 40 \\
        AY.44   & 1416   & AY.75 & 37 \\
        AY.100   & 717   & AY.3.1 & 30 \\
        AY.3   & 710   & AY.3.3 & 28 \\
        AY.25   & 585   & AY.107 & 27 \\
        AY.25.1   & 382   & AY.34.1 & 25  \\
        AY.39   & 248   & AY.46.6 & 21 \\
        AY.119   & 242   & AY.98.1  &  20 \\
        B.1.617.2   & 175   & AY.13 & 19 \\
        AY.20   & 130   & AY.116.1 & 18 \\
        AY.26   & 107   & AY.126 & 17 \\
        AY.4   & 100   & AY.114 & 15 \\
        AY.117   & 94   & AY.125 & 14 \\
        AY.113   & 94   & AY.34 & 14 \\
        AY.118   & 86   & AY.46.1 & 14 \\
        AY.43   & 85   & AY.92 & 13 \\
        AY.122   & 84   & AY.98 & 12 \\
        BA.1   & 79   & AY.46.4 & 12 \\
        AY.119.2   &  74   & AY.127 & 12 \\
        AY.47   & 73   & AY.111 & 10 \\
        AY.39.1   & 70   & \_ & \_ \\

      \hline
  \end{tabular}
  \label{tbl_variant_information}
\end{table}

\section{Appendix B: Data Visualization}\label{sec_data_viz}
\begin{figure}[h!]
    \centering
    \begin{subfigure}{.5\textwidth}
        \centering
        \includegraphics[scale = 0.4]{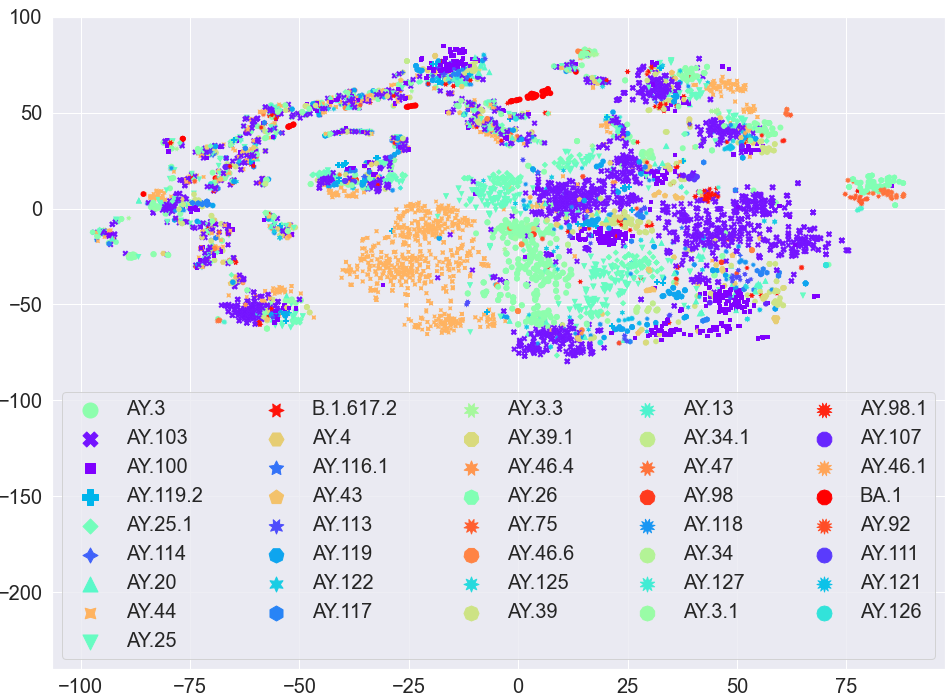}
        \caption{Spike2Vec}
        \label{fig_spike2vec_tsne}
    \end{subfigure}%
    \\
    \begin{subfigure}{.5\textwidth}
        \centering
        \includegraphics[scale = 0.4]{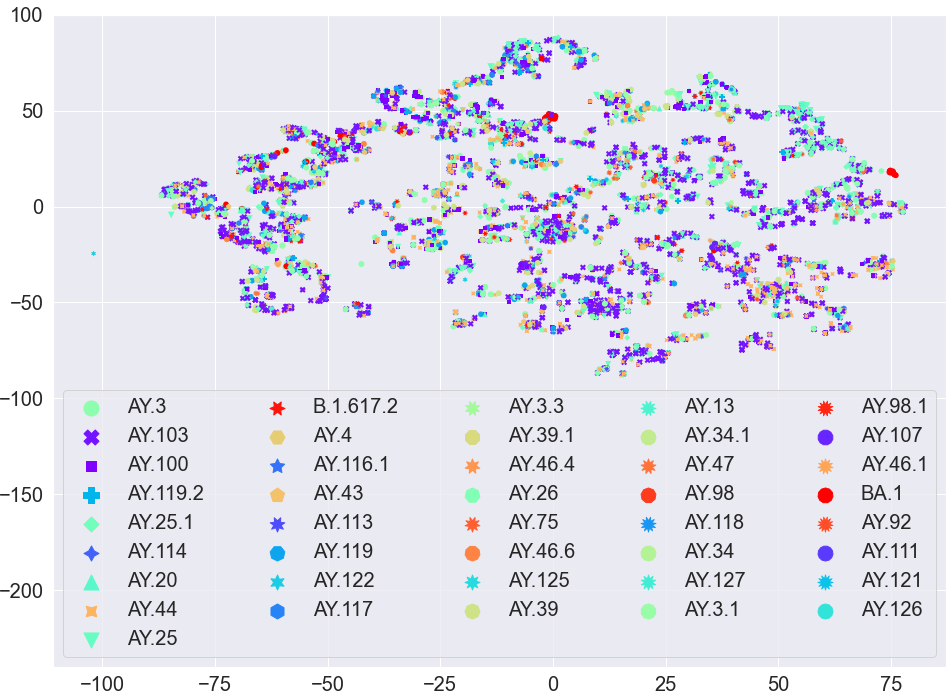}
    \caption{PSSM2Vec}
    \label{fig_pssm2vec_tsne}
    \end{subfigure}%
    \\
    \begin{subfigure}{.5\textwidth}
        \centering
        \includegraphics[scale = 0.4]{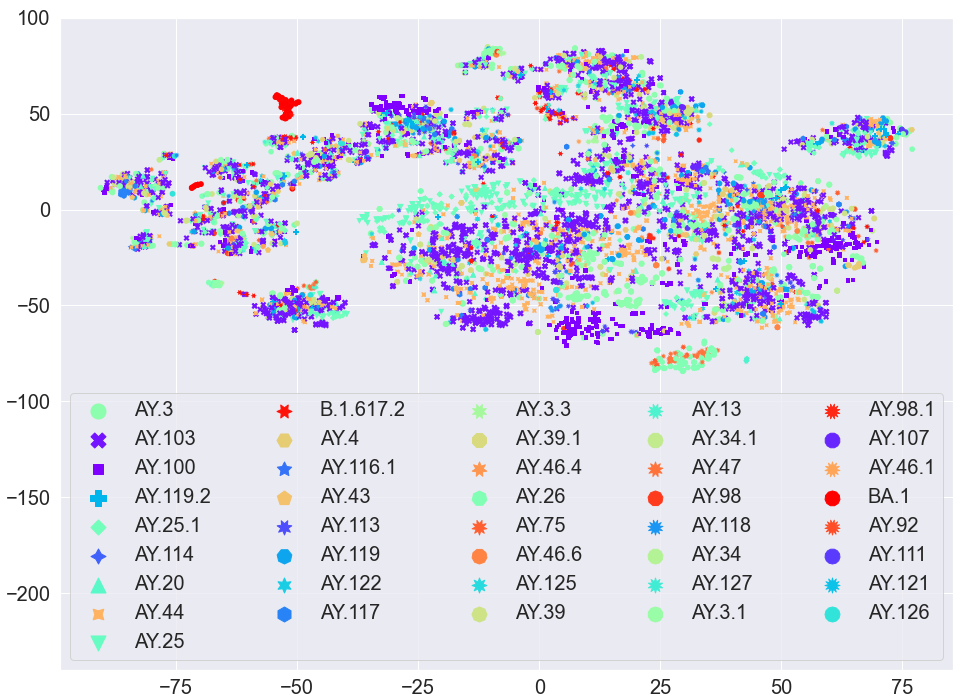}
    \caption{Min2Vec}
    \label{fig_min2vec_tsne}
    \end{subfigure}
    \caption{t-SNE plot for different embedding methods.}
    \label{fig_tsnes}
\end{figure}

To visualise if there is any (natural) clustering in our data, we generated 2d representation of the feature embeddings using t-distributed stochastic neighbor embedding (t-SNE) approach~\citep{van2008visualizing}. The main advantage of t-SNE is that it preserves the pair-wise distance between vectors in $2$ dimensions.
The t-SNE plot for different coronavirus variants is given in Figure~\ref{fig_spike2vec_tsne}, Figure~\ref{fig_pssm2vec_tsne}, and Figure~\ref{fig_min2vec_tsne} for Spike2Vec, PSSM2Vec, and Min2Vec, respectively. For Spike2Vec based t-SNE plot, we can observe that some of the variants (e.g. AY.103) is grouped more clearly than the other variants. PSSM2Vec however maintains smaller groups of variants as compared to Spike2Vec. The structure of Min2Vec based t-SNE is more similar to Spike2Vec, however, it grouped some other variants (e.g. AY.96.1) more clearly as compared to Spike2Vec. In general, we can observe that all embedding methods preserves the overall structure of the data.

\pagebreak
\section{Appendix C: Results and Discussion}
The standard deviation of $5$ runs for the original data (without any errored sequences) is given in Table~\ref{tbl_std_org}. Note the the accuracies (average values of $5$ runs) for the same data are reported in Table~\ref{tbl_acc_org} in the main paper.

\begin{table}[h!]
  \centering
  \caption{Standard Deviation Results on 8220 (original) nucleotide sequences (without any error).}
    \resizebox{0.95\textwidth}{!}{
  \begin{tabular}{p{1.9cm}p{1.3cm}p{1.3cm}p{1.3cm}p{1.3cm}p{1.3cm}cp{1.3cm} | p{1.3cm}}
    \hline
    \multirow{3}{1.1cm}{Embed. Method} & \multirow{3}{0.7cm}{ML Algo.} & \multirow{3}{*}{Acc.} & \multirow{3}{*}{Prec.} & \multirow{3}{*}{Recall} & \multirow{3}{0.9cm}{F1 weigh.} & \multirow{3}{0.9cm}{F1 Macro} & \multirow{3}{1.2cm}{ROC- AUC} & Train. runtime (sec.) \\	
    \hline	\hline	
    
    \multirow{7}{*}{Spike2Vec}  
    & SVM & 0.009337 & 0.008654 & 0.009337 & 0.009122 & 0.011612 & 0.011101 & 0.300614 \\
    & NB & 0.003307 & 0.059135 & 0.003307 & 0.003323 & 0.002726 & 0.007099 & 0.017888 \\
    & MLP & 0.012523 & 0.015917 & 0.012523 & 0.015363 & 0.028586 & 0.017015 & 3.836258 \\
    & KNN & 0.009530 & 0.011883 & 0.009530 & 0.010848 & 0.022703 & 0.011329 & 0.036290 \\
    & RF & 0.005541 & 0.006865 & 0.005541 & 0.006641 & 0.028757 & 0.014456 & 0.263474 \\
    & LR & 0.005504 & 0.004573 & 0.005504 & 0.005868 & 0.017053 & 0.012580 & 0.867295 \\
    & DT & 0.004102 & 0.002976 & 0.004102 & 0.003101 & 0.013446 & 0.010527 & 0.032883 \\

    \cline{2-9}	
    \multirow{7}{*}{PSSM2Vec}  
    & SVM & 0.007223 & 0.003972 & 0.007223 & 0.005558 & 0.000217 & 0.000000 & 0.118888 \\
    & NB & 0.002015 & 0.007148 & 0.002015 & 0.000983 & 0.004011 & 0.006194 & 0.008142 \\
    & MLP & 0.005716 & 0.015879 & 0.005716 & 0.006257 & 0.009698 & 0.005917 & 2.603367 \\
    & KNN & 0.009810 & 0.013064 & 0.009810 & 0.010757 & 0.023313 & 0.010837 & 0.028786 \\
    & RF & 0.006478 & 0.008212 & 0.006478 & 0.007973 & 0.017069 & 0.007942 & 0.052490 \\
    & LR & 0.007223 & 0.003972 & 0.007223 & 0.005558 & 0.000217 & 0.000000 & 0.025634 \\
    & DT & 0.006273 & 0.009446 & 0.006273 & 0.007321 & 0.013302 & 0.007477 & 0.009866 \\
    \cline{2-9}	
    \multirow{7}{*}{Min2Vec}  
    & SVM & 0.008510 & 0.008449 & 0.009629 & 0.008932 & 0.015522 & 0.010247 & 0.281588 \\
    & NB & 0.004464 & 0.060411 & 0.004464 & 0.006046 & 0.02021 & 0.007911 & 0.016351 \\
     & MLP & 0.011599 & 0.007398 & 0.011599 & 0.009882 & 0.028691 & 0.015814 & 1.480781 \\
     & KNN & 0.006601 & 0.009581 & 0.006601 & 0.007661 & 0.014181 & 0.005275 & 0.014744 \\
     & RF & 0.004837 & 0.004967 & 0.004837 & 0.006467 & 0.034441 & 0.017324 & 0.044175 \\
     & LR & 0.001902 & 0.005262 & 0.001902 & 0.002995 & 0.014008 & 0.004733 & 0.261828 \\
     & DT & 0.010011 & 0.011210 & 0.010011 & 0.010262 & 0.024173 & 0.014445 & 0.011970 \\

    \hline
  \end{tabular}
  }
  \label{tbl_std_org}
\end{table}

\subsection{Robustness Results on Errored Sequences}
The robustness results of PBSIM based errored sequences (with depth $15$ and $20$) are shown in Table~\ref{tbl_pbsim_depth_15_and_20}. 
Similarly the robustness results with Illumina based errored sequences having number of short reads as $15000$ and $20000$ are shown in Table~\ref{tbl_illumina_15k_20k_reads}.

\begin{table}[h!]
  \centering
  \caption{Robustness Results on PBSIM data with $15$ and $20$ as read depth.}
    \resizebox{0.95\textwidth}{!}{
  \begin{tabular}{@{\extracolsep{4pt}}p{1.9cm}p{0.5cm}p{0.4cm}p{0.4cm}p{0.7cm}p{0.7cm}cp{0.7cm} p{1.1cm} p{0.4cm}p{0.4cm}p{0.7cm}p{0.7cm}cp{0.7cm} p{1.1cm}}
    \hline
    & & \multicolumn{7}{c}{Depth: 15} & \multicolumn{7}{c}{Depth: 20}
    \\
    \cline{3-9} \cline{10-16}
    \multirow{3}{1.1cm}{Embed. Method} & \multirow{3}{0.7cm}{ML Algo.} & \multirow{3}{*}{Acc.} & \multirow{3}{*}{Prec.} & \multirow{3}{*}{Recall} & \multirow{3}{0.9cm}{F1 weigh.} & \multirow{3}{0.9cm}{F1 Macro} & \multirow{3}{1.2cm}{ROC- AUC} & Train. runtime (sec.) & \multirow{3}{*}{Acc.} & \multirow{3}{*}{Prec.} & \multirow{3}{*}{Recall} & \multirow{3}{0.9cm}{F1 weigh.} & \multirow{3}{0.9cm}{F1 Macro} & \multirow{3}{1.2cm}{ROC- AUC} & Train. runtime (sec.) \\	
    \hline	\hline	
    \multirow{7}{*}{Spike2Vec}  
     & SVM & 0.01 & 0.00 & 0.01 & 0.00 & 0.00 & 0.50 & 11.19 & 0.01 & 0.00 & 0.01 & 0.00 & 0.00 & 0.50 & 11.17 \\
 & NB & 0.00 & 0.00 & 0.00 & 0.00 & 0.00 & 0.50 & 0.81 & 0.00 & 0.00 & 0.00 & 0.00 & 0.00 & 0.50 & 0.70 \\
 & MLP & 0.00 & 0.00 & 0.00 & 0.00 & 0.00 & 0.50 & 21.66 & 0.01 & 0.00 & 0.01 & 0.00 & 0.00 & 0.50 & 20.82 \\
 & KNN & 0.28 & 0.08 & 0.28 & 0.12 & 0.01 & 0.50 & 2.32 & 0.28 & 0.08 & 0.28 & 0.12 & 0.01 & 0.50 & 2.24 \\
 & RF & 0.28 & 0.08 & 0.28 & 0.12 & 0.01 & 0.50 & 2.48 & 0.28 & 0.08 & 0.28 & 0.12 & 0.01 & 0.50 & 2.42 \\
 & LR & 0.01 & 0.00 & 0.01 & 0.00 & 0.00 & 0.50 & 11.89 & 0.01 & 0.00 & 0.01 & 0.00 & 0.00 & 0.50 & 11.77 \\
 & DT & 0.01 & 0.00 & 0.01 & 0.00 & 0.00 & 0.50 & 0.31 & 0.01 & 0.00 & 0.01 & 0.00 & 0.00 & 0.50 & 0.31 \\
    \cline{2-9}	\cline{10-16}	
    \multirow{7}{*}{PSSM2Vec}  
     & SVM & 0.28 & 0.08 & 0.28 & 0.12 & 0.01 & 0.50 & 9.97 & 0.28 & 0.08 & 0.28 & 0.12 & 0.01 & 0.50 & 9.54 \\
 & NB & 0.01 & 0.00 & 0.01 & 0.00 & 0.00 & 0.50 & 0.15 & 0.01 & 0.00 & 0.01 & 0.00 & 0.00 & 0.50 & 0.16 \\
 & MLP & 0.01 & 0.00 & 0.01 & 0.00 & 0.00 & 0.50 & 15.63 & 0.01 & 0.00 & 0.01 & 0.00 & 0.00 & 0.50 & 19.20 \\
 & KNN & 0.01 & 0.00 & 0.01 & 0.00 & 0.00 & 0.50 & 2.21 & 0.01 & 0.00 & 0.01 & 0.00 & 0.00 & 0.50 & 2.18 \\
 & RF & 0.00 & 0.00 & 0.00 & 0.00 & 0.00 & 0.50 & 2.03 & 0.00 & 0.00 & 0.00 & 0.00 & 0.00 & 0.50 & 1.94 \\
 & LR & 0.28 & 0.08 & 0.28 & 0.12 & 0.01 & 0.50 & 1.13 & 0.28 & 0.08 & 0.28 & 0.12 & 0.01 & 0.50 & 1.11 \\
 & DT & 0.00 & 0.00 & 0.00 & 0.00 & 0.00 & 0.50 & 0.09 & 0.00 & 0.00 & 0.00 & 0.00 & 0.00 & 0.50 & 0.08 \\
    \cline{2-9}	\cline{10-16}	
    \multirow{7}{*}{Min2Vec}  
     & SVM & 0.01 & 0.01 & 0.01 & 0.00 & 0.01 & 0.50 & 15.47 & 0.01 & 0.01 & 0.01 & 0.00 & 0.01 & 0.50 & 17.54 \\
 & NB & 0.00 & 0.00 & 0.00 & 0.00 & 0.00 & 0.50 & 0.77 & 0.00 & 0.00 & 0.00 & 0.00 & 0.00 & 0.50 & 0.84 \\
 & MLP & 0.05 & 0.00 & 0.05 & 0.00 & 0.00 & 0.50 & 24.19 & 0.05 & 0.00 & 0.05 & 0.00 & 0.00 & 0.50 & 34.86 \\
 & KNN & 0.05 & 0.00 & 0.05 & 0.00 & 0.00 & 0.50 & 2.50 & 0.05 & 0.00 & 0.05 & 0.00 & 0.00 & 0.50 & 1.93 \\
 & RF & 0.09 & 0.01 & 0.09 & 0.01 & 0.00 & 0.50 & 2.39 & 0.28 & 0.08 & 0.28 & 0.12 & 0.01 & 0.50 & 2.25 \\
 & LR & 0.00 & 0.00 & 0.00 & 0.00 & 0.00 & 0.50 & 9.45 & 0.00 & 0.00 & 0.00 & 0.00 & 0.00 & 0.50 & 10.11 \\
 & DT & 0.07 & 0.01 & 0.07 & 0.01 & 0.00 & 0.50 & 0.20 & 0.07 & 0.01 & 0.07 & 0.01 & 0.00 & 0.50 & 0.19 \\
    \hline
  \end{tabular}
  }
  \label{tbl_pbsim_depth_15_and_20}
\end{table}

\begin{table}[h!]
  \centering
   \caption{Robustness Results on Illumina based errored sequences with $15000$ and $20000$ short reads used in the simulation process.}
    \resizebox{0.95\textwidth}{!}{
  \begin{tabular}{@{\extracolsep{4pt}}p{1.9cm}p{0.5cm}p{0.4cm}p{0.4cm}p{0.7cm}p{0.7cm}cp{0.7cm} p{1.1cm} p{0.4cm}p{0.4cm}p{0.7cm}p{0.7cm}cp{0.7cm} p{1.1cm}}
    \hline
    & & \multicolumn{7}{c}{\# of Short Reads: 15000} & \multicolumn{7}{c}{\# of Short Reads: 20000}
    \\
    \cline{3-9} \cline{10-16}
    \multirow{3}{1.1cm}{Embed. Method} & \multirow{3}{0.7cm}{ML Algo.} & \multirow{3}{*}{Acc.} & \multirow{3}{*}{Prec.} & \multirow{3}{*}{Recall} & \multirow{3}{0.9cm}{F1 weigh.} & \multirow{3}{0.9cm}{F1 Macro} & \multirow{3}{1.2cm}{ROC- AUC} & Train. runtime (sec.) & \multirow{3}{*}{Acc.} & \multirow{3}{*}{Prec.} & \multirow{3}{*}{Recall} & \multirow{3}{0.9cm}{F1 weigh.} & \multirow{3}{0.9cm}{F1 Macro} & \multirow{3}{1.2cm}{ROC- AUC} & Train. runtime (sec.) \\	
    \hline	\hline	
    \multirow{7}{*}{Spike2Vec}  
     & SVM & 0.68 & 0.69 & 0.68 & 0.67 & 0.44 & 0.72 & 12.41 & 0.68 & 0.69 & 0.68 & 0.67 & 0.44 & 0.72 & 11.60 \\
 & NB & 0.00 & 0.00 & 0.00 & 0.00 & 0.03 & 0.52 & 0.95 & 0.00 & 0.00 & 0.00 & 0.00 & 0.03 & 0.52 & 0.65 \\
 & MLP & 0.63 & 0.64 & 0.63 & 0.62 & 0.32 & 0.67 & 26.30 & 0.63 & 0.65 & 0.63 & 0.63 & 0.31 & 0.66 & 20.25 \\
 & KNN & 0.51 & 0.50 & 0.51 & 0.45 & 0.13 & 0.56 & 2.50 & 0.51 & 0.50 & 0.51 & 0.45 & 0.13 & 0.56 & 2.29 \\
 & RF & 0.71 & 0.73 & 0.71 & 0.68 & 0.51 & 0.71 & 3.17 & 0.71 & 0.72 & 0.71 & 0.68 & 0.49 & 0.71 & 2.64 \\
 & LR & 0.71 & 0.70 & 0.71 & 0.69 & 0.47 & 0.73 & 12.22 & 0.71 & 0.70 & 0.71 & 0.69 & 0.47 & 0.73 & 12.22 \\
 & DT & 0.53 & 0.55 & 0.53 & 0.53 & 0.35 & 0.68 & 0.33 & 0.52 & 0.54 & 0.52 & 0.52 & 0.31 & 0.66 & 0.32 \\
    \cline{2-9}	\cline{10-16}	
    \multirow{7}{*}{PSSM2Vec}  
     & SVM & 0.28 & 0.08 & 0.28 & 0.12 & 0.01 & 0.50 & 9.79 & 0.28 & 0.08 & 0.28 & 0.12 & 0.01 & 0.50 & 9.59 \\
 & NB & 0.00 & 0.00 & 0.00 & 0.00 & 0.00 & 0.50 & 0.18 & 0.00 & 0.00 & 0.00 & 0.00 & 0.00 & 0.50 & 0.22 \\
 & MLP & 0.22 & 0.20 & 0.22 & 0.14 & 0.03 & 0.51 & 15.89 & 0.18 & 0.18 & 0.18 & 0.14 & 0.02 & 0.51 & 19.92 \\
 & KNN & 0.17 & 0.22 & 0.17 & 0.14 & 0.02 & 0.51 & 1.98 & 0.17 & 0.22 & 0.17 & 0.14 & 0.02 & 0.51 & 2.28 \\
 & RF & 0.12 & 0.17 & 0.12 & 0.12 & 0.03 & 0.51 & 1.76 & 0.13 & 0.18 & 0.13 & 0.13 & 0.03 & 0.51 & 2.00 \\
 & LR & 0.28 & 0.08 & 0.28 & 0.12 & 0.01 & 0.50 & 1.01 & 0.28 & 0.08 & 0.28 & 0.12 & 0.01 & 0.50 & 1.03 \\
 & DT & 0.13 & 0.19 & 0.13 & 0.13 & 0.03 & 0.51 & 0.09 & 0.12 & 0.18 & 0.12 & 0.12 & 0.03 & 0.51 & 0.08 \\
    \cline{2-9}	\cline{10-16}	
    \multirow{7}{*}{Min2Vec}  
    & SVM & 0.52 & 0.53 & 0.52 & 0.48 & 0.31 & 0.65 & 19.16 & 0.52 & 0.54 & 0.52 & 0.48 & 0.32 & 0.65 & 18.63 \\
 & NB & 0.02 & 0.07 & 0.02 & 0.01 & 0.06 & 0.55 & 0.95 & 0.02 & 0.07 & 0.02 & 0.01 & 0.06 & 0.55 & 0.77 \\
 & MLP & 0.50 & 0.45 & 0.50 & 0.45 & 0.22 & 0.59 & 33.36 & 0.52 & 0.48 & 0.52 & 0.48 & 0.29 & 0.63 & 37.64 \\
 & KNN & 0.37 & 0.35 & 0.37 & 0.31 & 0.10 & 0.55 & 2.42 & 0.37 & 0.35 & 0.37 & 0.31 & 0.10 & 0.55 & 2.34 \\
 & RF & 0.64 & 0.68 & 0.64 & 0.60 & 0.46 & 0.69 & 2.21 & 0.64 & 0.67 & 0.64 & 0.60 & 0.45 & 0.69 & 2.44 \\
 & LR & 0.51 & 0.53 & 0.51 & 0.46 & 0.30 & 0.64 & 9.14 & 0.51 & 0.53 & 0.51 & 0.46 & 0.30 & 0.64 & 9.38 \\
 & DT & 0.51 & 0.53 & 0.51 & 0.51 & 0.35 & 0.68 & 0.19 & 0.51 & 0.53 & 0.51 & 0.51 & 0.36 & 0.68 & 0.19 \\
    \hline
  \end{tabular}
  }
  \label{tbl_illumina_15k_20k_reads}
\end{table}


\end{document}